\documentstyle[pra,floats,eqsecnum,aps]{revtex}

\begin{document} \draft

\title{
Dual metrics and non-generic supersymmetries for a class of Siklos
spacetimes}

\author{D. Baleanu \footnote{electronic address:
dumitru@cankaya.edu.tr}}

\address{
Department of Mathematics and Computer Sciences, Faculty of Arts and
Sciences, {\c C}ankaya University, 06531 Ankara, Turkey\\
\footnote{On leave of absence}Institute of Space Sciences, P. O. Box,
MG-23, R-76900, Magurele-Bucharest, Romania}

\author{S. Ba{\c s}kal \footnote{electronic address:
baskal@newton.physics.metu.edu.tr}}

\address{
Department of Physics, University of Maryland, College Park, MD 20742,
USA\\
\footnote{On leave of absence} Department of Physics, Middle East
Technical University, 06531 Ankara, Turkey}

\maketitle

\begin{abstract}
The presence of Killing-Yano tensors implies the existence of 
non-generic supercharges in spinning point particle theories 
on curved backgrounds.  Dual metrics are defined through their 
associated non-degenerate Killing tensors of valence two. 
Siklos spacetimes, which are the only non-trivial Einstein spaces
conformal to non-flat pp-waves are investigated in regards to the
existence of their corresponding Killing and Killing-Yano tensors.
It is found that under some restrictions, pp-wave metrics and Siklos
spacetimes admit dual metrics and non-generic supercharges.  Possible
significance of those dual spacetimes are discussed.
\end{abstract}



\section{Introduction}
The most interesting feature of Killing-Yano (KY) tensors \cite{yano52} 
is established
in the context of pseudo-classical spinning point particles, with N=1 
world line supersymmetry, as an object generating supercharges that depend
on the background metric \cite{berezin77},
\cite{gibbons93},\cite{holten95}, 
\cite{holten99}, \cite{visinescu01},\cite{tanimoto95}.
More intriguingly, the separability of the Dirac equation 
in the Kerr geometry is traced back to the existence of KY
tensor, in that background \cite{carter68},\cite{carter79}.
On the other hand Killing tensors, in some cases can be considered
the "square" of KY tensors.  They give rise to the associated constant of
motion and play an important role in the complete solution of the
geodesic equation. Furthermore, it has been shown that there is a
reciprocal relation between spaces admitting non-degenerate Killing
tensors of valence two and  the so-called "dual" spaces whose metrics are
specified through those Killing tensors \cite{holten96}.  
Such novel aspects promoted literature on Killing and KY tensors, 
which have long known to relativists \cite{tachibana68},
\cite{collinson74},  \cite{jezierski97},\cite{hall87}, \cite{kramer80}. 
Recently, the Killing spacetimes for St{\" a}ckel systems of three
dimensional separable coordinates were 
investigated \cite{hinterleitner99}. 
More recently, further generalizations of Killing tensors and
their existence criteria were discussed \cite{howarth00}. 
Non-degenerate Killing tensors of valence two were investigated for
a class of metrics describing pure radiative spacetimes \cite{baleanu01}.
Lax tensors \cite{rosquist97}, in some specific cases are shown to be 
Killing-Yano tensors of order three \cite{baleanu00}. 
Static axisymmetric spacetimes \cite{tanimoto94} and the Euclidean
Taub-NUT metric \cite{visinescu01} were
analyzed within the framework of pseudo-classical spinning particles,
as well as the spacetime manifolds with constant
curvature \cite{baleanu99},\cite{goenner01}. 

The purpose of this paper is twofold. First, we believe that a
systematical investigation of spacetimes admitting Killing-Yano 
tensor will strengthen the connection between the background geometry 
and the non-generic supersymmetries, especially in such cases when the
background metrics receive significant physical interpretations.
We also know that properly contracted KY tensors generate a constant 
of motion by defining a Killing tensor. 
Furthermore, those Killing tensors, as well as those that are solved
through the defining equations describe dual spacetime, on the
condition that they are non-degenerate.  On the other hand,
finding non-degenerate Killing tensors on a particular
spacetime, is not an easy task, because the condition of
non-degeneracy is imposed by hand and has no connection with
symmetries of the equations.  Therefore, providing examples of 
physical significance is a step further towards a better 
understanding of dual spacetimes. 

Secondly, the spacetimes that we are going to investigate from the above
point of view, have been of great interest since they were first
introduced.  The metric describing plane fronted waves with parallel rays
(pp-wave) is very well-known \cite{kramer80}.  On the other hand Siklos
spacetimes are the only non-trivial Einstein spaces conformal to non-flat
pp-waves \cite{siklos85}. In fact, the initial motivation to study such
spacetimes was that they occurred naturally in $N=1$ supergravity. 
The presence of a negative cosmological 
constant in the Siklos metric implies that the space is not
asymptotically flat, a probable feature of how our real universe 
may be.  It has been shown that a through analysis of particle
motion in Siklos spacetimes require rotating reference 
frames \cite{podolskygriff98},\cite{mashhoon00}.  
Therefore, the importance of constants of motion in such spacetimes 
can hardly be overestimated.
Because of those reasons we believe that a systematical
investigation of Killing and Killing-Yano tensors of pp-waves and Siklos
type spacetimes is worthwhile.

We will carry out our investigation according to the following strategy:
We will solve the KY equation for valence two and three
and classify the metrics.  We will optimize between having
a less restrictive metric and attaining the maximum number of
non-vanishing KY components.  We will exclude all flat background
solutions.  From those KY tensors, of valence two and three, it is 
straightforward to calculate the associated Killing tensors, and the 
dual metrics.  However, those are not the only Killing tensors on the
manifold.  We will also solve the Killing equations.  Since  
we are interested in dual spaces, we will look for Killing tensors having 
the same surviving components as that of the original metric.

In the following section, we will give the basic formulations.
We will also discuss the number of independent equations to be solved
for KY tensors.  In the third section we will classify the pp-wave and
the Siklos metric, so that they admit Killing and KY tensors of valence
two and three.  A class of Killing and KY tensor solutions will
be written explicitly.  Associated dual spacetimes will also be
calculated.  The last section will be devoted to concluding remarks. 

\section{Killing and Killing-Yano tensors}
\subsection{Killing-Yano tensors and non-generic supercharges}
A Killing-Yano tensor of valence n, $f_{\nu_{1}\nu{2}\cdots\nu_{n}}$,
is an antisymmetric tensor fulfilling the following equations:
\begin{equation}
{\cal M}_{\lambda\,\nu_{1}\nu_{2}\cdots\nu_{n}}   
\, \equiv \, f_{\nu_{1}\nu_{2}\cdots(\nu_{n};\lambda)}\,=\,0,
\label{kye}
\end{equation}
where semicolon denotes the covariant derivative.

Since we shall be investigating the solutions of (\ref{kye}) 
for $n=2$ and for $n=3$, in four dimensions,
it is worthwhile to discuss the number of independent equations.
For $n=2$, the number of independent
equations can be found by analyzing the symmetries
of ${\cal M}_{\mu\nu\alpha}$, in regards to its repeated and 
all-distinct indices: If all the indices are repeated
then ${\cal M}_{\alpha\alpha\alpha}$ is identically zero.
If two indices are repeated then the symmetry is
${\cal M}_{\alpha\nu\nu}=-{\cal M}_{\nu\alpha\nu}=
-{\cal M}_{\nu\nu\alpha}$, yielding 12 independent components.
If all the indices are distinct then 
${\cal M}_{\mu\nu\alpha}={\cal M}_{\nu\mu\alpha}$, resulting in
another 12 independent components. 
Therefore, there are totally 24 independent equations, to
be solved for the six independent components of the 
Killing-Yano tensor of rank two.

Similar arguments are also valid for Killing-Yano tensors of valence
three. If all the indices or three of the indices are the same
then the equations are identically zero.  If two are the same then
$\cal M_{\alpha\alpha\mu\nu}=- \cal M_{\alpha\alpha\nu\mu}$, and there 
are 12 independent equations. If all are distinct then
$\cal M_{\alpha\beta\mu\nu}=\cal M_{\beta\alpha\mu\nu}=
-\cal M_{\alpha\beta\nu\mu}$ and so
there are 4 more equations, all sum up to 16 independent equations
to be solved for four independent components.

The spinning particle model was constructed to be
supersymmetric \cite{gibbons93}, 
therefore independent of the form of the metric
there is always a conserved supercharge $Q_{0}=\Pi_{\mu}\psi^{\mu}$.
Here, $\Pi_{\mu}$ is the covariant momenta and $\psi^{\mu}$ are odd
Grassmann variables.
The existence of Killing-Yano of valence $n$ are related to 
non-generic supersymmetries described by the following supercharge
\begin{equation}
Q_{f}=f_{\nu_{1}\nu_{2}\cdots\nu_{r}}\Pi^{\nu_{1}}\psi^{\nu_{2}}\cdots
\psi^{\nu_{r}}
\label{sc}
\end{equation}
which is a superinvariant: $\{Q_{0}, Q_{f}\}=0$.
The Jacobi identities and (\ref{kye}) guarantee that it is
also a constant of motion: $\{Q_{f},H\}=0$, with
\begin{equation}
H=\frac{1}{2}g^{\mu\nu}\Pi_{\mu}\Pi_{\nu}
\label{eyc}
\end{equation}
and with the appropriate definitions of the brackets.

\subsection{Killing tensors and dual spacetimes}
A Killing tensor of valence two is defined through the equation
\begin{equation}
K_{(\mu\nu;\alpha)}=0.
\label{kte}
\end{equation}

It has been shown in detail in reference \cite{holten96} that $K^{\mu\nu}$
and $g^{\mu\nu}$ are reciprocally the contravariant components of the
Killing tensors with respect to each other.

If $K^{\mu\nu}$ is non-degenerate, then through the relation
\begin{equation}
K^{\mu \alpha}k_{\alpha \nu}=\delta^{\mu}\,_{\nu},
\label{dual}
\end{equation}
the second rank non-degenerate tensor $k_{\mu\nu}$, 
can be viewed as the metric on the "dual" space.
The relation between the Christoffel symbols,
$\hat\Gamma^{\mu}\,_{\alpha \beta}$ of the dual and of the initial
manifold
can be expressed by writing $\hat\Gamma^{\mu}\,_{\alpha \beta}$ in terms 
of the Killing tensor and taking (\ref{kte}) into 
account~\cite{baleanu01}:
\begin{equation}
\hat \Gamma^{\mu}\,_{\alpha \beta} = \Gamma^{\mu}\,_{\alpha \beta}
-{\cal K}^{\mu\delta}K_{\alpha\beta;\delta},
\end{equation}
where ${\cal K}^{\mu \alpha}K_{\alpha \nu}=\delta^{\mu}\,_{\nu}$.

The notion of geometric duality extends to that of phase space.
The constant of motion
$K=\frac{1}{2}K^{\mu\nu}\Pi_{\mu}\Pi_{\nu}$,
generates symmetry transformations on the phase space linear
in momentum: $\{ x^{\mu},K \}=K^{\mu \nu}\Pi_{\nu}$, and in view of
(\ref{kte}) the Poisson brackets satisfy $\{ H,K \}=0$,
where $H$ is as in (\ref{eyc}).  Thus, in the 
phase space there is a reciprocal model with constant of motion $H$ 
and the Hamiltonian $K$.

Killing-Yano tensors of any valence can be considered as the square root 
of the Killing tensors of valence two in the sense that, their appropriate
contractions yield
\begin{equation}
K_{\mu\nu}=g^{\alpha\beta}f_{\mu\alpha}f_{\beta\nu}
\label{kfky1}
\end{equation}
or for valence three it can be written as
\begin{equation}
K_{\mu\nu}=g^{\alpha\delta}g^{\beta\gamma}
f_{\mu\alpha\beta}f_{\gamma\delta\nu}.
\label{kfky2}
\end{equation}

\section{The pp-wave metric and Siklos spacetimes}
\subsection{The pp-wave metric}
The very well-known pp-wave metric, describes plane fronted waves with
parallel rays, admit a non-expanding shear-free and twist-free null-
geodesic congruence, can be expressed in the form \cite{kundt61}:
\begin{equation}
ds^2= 2dudv+dx^2 +dy^2 +h(x,y,u)du^2.
\label{pp}
\end{equation}
Here, $x^{\mu}=(v,x,y,v)$, $x$ and $y$ are spacial coordinates, 
$u$ is the retarded time.  The Killing vector  $l^{\mu}=\delta^{\mu}_{1}$
is at the same time tangent to the null geodesics, so the coordinate
$v$ can be considered as an affine parameter.
Many of its characteristics have been exploited in the 
literature for various purposes \cite{podolsky98},\cite{baskal99}. 

\subsubsection{Subclasses admitting Killing-Yano tensors}
Here, we shall investigate the Killing-Yano
tensors of the pp-wave metric.  We will search for 
KY tensors with maximum number of components, with minimum 
restrictions on the metric function.  Here and thereafter, we have
excluded all flat solutions.
We shall spare the reader from all computational details, and give 
the results.  Incidentally, it may be interesting to note that,
we can have Killing-Yano tensors with no restrictions on the metric
as in the following:  A two-component KY tensor
\begin{equation}
f_{24}=c_{1}, \qquad
f_{34}=c_{2}
\label{ky2pg}
\end{equation}
exist for any form of $h(x,y,u)$.  However, it is apparent that this
tensor is trivial ($f_{\mu\nu;\alpha}=0$).  

This metric admits a KY tensor with at most four non-zero components: 
\begin{equation}
\begin{array}{lll}
f_{12}=0, & f_{13}=0, & f_{14}=c_{1},\\ [2mm]
f_{23}=c_{2},& f_{24}=r(u), & f_{34}=s(u).
\label{ky2p}
\end{array}
\end{equation}
with the following restrictions on the metric function: 
\begin{equation}
\begin{array}{l}
2r(u)_{,u} - c_{1} h_{,x}  + c_{2} h_{,y} = 0, \\[2mm]
2s(u)_{,u} - c_{1} h_{,y} - c_{2} h_{,x} =0.
\label{c2}
\end{array}
\end{equation}
Analyzing the integrability conditions it is found that, for 
pp-wave metric there is no solution for non-zero
$f_{12}$ and/or $f_{13}$ components, even if one nullifies
some of or all of the other components.

A KY tensor of order three has three non-vanishing components:
\begin{equation}
\begin{array}{ll}
f_{123}=0, & f_{124}=c_{1},\\
f_{134}=c_{2}, & f_{234}=q(u)
\end{array}
\label{ky3p}
\end{equation}
where $q(u)$ and the metric function are subject to: 
\begin{equation}
2q(u)_{,u} - c_{2} h(x,y,u)_{,x} + c_{1} h(x,y,u)_{,y}=0.
\label{c3}
\end{equation}
Similar arguments as above also apply here about the vanishing of
$f_{123}$.

\subsubsection{Killing tensors}
Recently,  Killing tensors for pp-wave metric are presented within the
framework of dual spacetimes, by solving (\ref{kte}) \cite{baleanu01}.  
It is found that the associated dual metrics 
fall into a class of spacetimes describing parallel null one-planes
given by Walker \cite{walker50},\cite{oktem76}.  It is possible to
associate a massless particle, with its four-momentum vector to be the
basis of the null one-plane.  If $\tilde{l}_{\mu}$ is the basis
of the plane, then $\tilde{l}_{\mu;\nu}=\alpha_{\nu}\tilde{l}_{\mu}$,
where $\alpha_{\nu}$ is the recurrence vector of the plane. 
Here we shall not repeat the same calculations; instead we will give the 
Killing tensors that can be obtained from KY tensors.
From (\ref{kfky1}) and (\ref{ky2p}) they become:
\begin{equation}
\begin{array}{l}
K_{14}=c_{1}^2,\qquad  K_{22}=-c_{2}^{2}, \qquad
K_{33}=-c_{2}^{2} \\[2mm] 
K_{24}=c_{1}r(u)+c_{2}s(u),  \\[2mm]
K_{34}=c_{1}s(u)-c_{2}r(u), \\[2mm] 
K_{44}=c_{1}^{2}\,h(x,y,u)-r(u)^{2}-s(u)^{2}. 
\end{array}
\label{kt2p}
\end{equation}
Dual metric obtained from above is:
\begin{equation}
\begin{array}{l}
k_{14}=1/c_{1}^{2}, \qquad  k_{22}=-1/2c_{2},
\qquad k_{33}=-1/c_{2}^{2},\\[2mm]
k_{24}=(c_{1}r(u)+c_{2}s(u))/2c_{1}^{2}c_{2},\\[2mm] 
k_{34}=(c_{1}s(u)-c_{2}r(u))/c_{1}^{2}c_{2}^{2},\\[2mm] 
k_{44}=\{2c_{2}^{2}(r(u)^{2}+s(u)^{2}+c_{1}^{2}h(x,y,u)
-2(c_{1}s(u)-c_{2}r(u))^{2}-c_{2}(c_{1}r(u)+c_{2}s(u))^{2}\}
/2c_{1}^{4}c_{2}^{2}.
\end{array}
\label{dp}
\end{equation}
With $c_{1}=1$, and $c_{2} \neq 0$, this metric also falls into 
a class of Walker's metric. 

Killing tensors obtained from (\ref{kfky2}) and 
(\ref{ky3p}) are:
\begin{equation}
\begin{array}{ll}
K_{14}=-2(c_{2}^{2}+c_{2}^{2}), \qquad & K_{22}=-2c_{1}^{2}, \\[2mm]
K_{23}=-2c_{1}c_{2}, \qquad & K_{24}=-2c_{2}q(u),\\[2mm]
K_{33}-2c_{2}^{2}, \qquad  & K_{34}=2c_{1}q(u), \\[2mm]
K_{44}=2(q(u)^{2}-(c_{1}^{2}+c_{2}^{2})h(x,y,u))
\end{array}
\label{kt3p}
\end{equation}
A straightforward calculation shows that,  dual metric cannot be 
obtained from this tensor, because its contravariant components 
turn out to form a singular matrix.

Killing tensors obtained from (\ref{ky2pg}) has only one
component: 
\begin{equation}
K_{44}=c_{1}^2+c_{2}^2,
\end{equation}
and cannot be considered as a metric for the
dual manifold, as (\ref{kt3p}) is.

\subsection{The Siklos metric}
The Siklos metric is expressed as
\begin{equation}
ds^{2}=\frac{\beta^{2}}{x^{2}}\left[2\,du\,dv +
dx^{2}+dy^{2}+h(x,y,u)\,du^{2}
\right]
\label{spp}
\end{equation}
where $\beta=\sqrt{-3/\Lambda}$ and $\Lambda$ is the negative
cosmological constant.  The presence of a negative cosmological 
constant implies that the spacetime is not asymptotically
flat.  It is demonstrated in \cite{siklos85}, that they represent
the only non-trivial Einstein spaces conformal to non-flat pp-waves.
As in pp-wave metric here also $u$ is the retarded time and the principal
null vector can be expressed as $l^{\mu}=\delta^{\mu}_{4}$.
Similar to that of the pp-wave metric the Siklos
metric also has vanishing optical parameters.  

\subsubsection{Subclasses admitting Killing-Yano tensors}
Thr Siklos metric admits a second order Killing-Yano tensor with
four non-zero components:
\begin{equation}
\begin{array}{ll}
f_{12}=0, \qquad & f_{13}=0 \\
f_{14}= (2r(u) - c \, y)/2x^{3}, \qquad & f_{23}= - c/2x^{2}\\
f_{24} = r(u)_{,u}/x^{2}, &
f_{34}= (y\,r(u)_{,u} + s(u) + c\,v)/x^{3}
\end{array}
\label{ky2s}
\end{equation}
where the functions $r(u)$ and $s(u)$ are related to the metric
function through the following differential equations:
\begin{equation}
\begin{array}{l}
(c_{1} y - 2 r(u)) \, h_{,x} - c_{1} x \,h_{,y}  + 4 x \,r(u)_{,uu}=0, \\ 
c_{1} x \, h_{,x} + (c_{1} y  - 2 r(u))\,  h_{,y}  - 2 c_{1}\, h + 4 y\,
r(u)_{,uu}+ 4 s(u)_{,u}=0,
\end{array}
\label{c4}
\end{equation}
with $h$ depending on all of its arguments.
A particular solution to these equations are found as:
\begin{equation}
\frac{r(u)_{,uu}}{r(u)}=h1(u), \qquad s(u)=c_{2}
\label{rs}
\end{equation}
where $c,\,c_{1}$ and $c_{2}$ are arbitrary constants.
Then, the metric function is found to be 
\begin{equation}
h=h1(u)(x^{2}+y^{2}).
\label{hs}
\end{equation}
Here, also as in the case of the pp-wave metric, there is no solution for
non-zero $f_{12}$ and/or $f_{13}$ components, due
to the integrability conditions.

The solutions for the third rank Killing-Yano tensor has only three
non-vanishing components:
\begin{equation}
\begin{array}{ll}
f_{123}=0,  & f_{124}=1/x^{3},\\[1mm]
f_{134}=(r(u)+y)/x^{4},\qquad &
f_{234}=r(u)_{,u}/x^{3}.
\end{array}
\label{ky3s}
\end{equation}
The metric function $h(x,y,u)$ and $r(u)$ are subject to the solutions of:
\begin{equation}
2x\,r(u)_{,uu} - r(u)\,h_{,x} - y \, h_{,x} + x\, h_{,y}\,=\,0.
\label{c5}
\end{equation}
We are led to assume the form of the metric as: $h=h3(u)[h1(x)+h2(y)]$.
Then the metric function becomes
\begin{equation}
h(x,y,u)=h3(u)(x^{2}-y^{2}),
\label{h2s}
\end{equation}
where $h3(u)$ can be found through
\begin{equation}
h3(u)=r(u)_{,uu}/ r(u)
\label{c6}.
\end{equation}

\subsubsection{Subclasses admitting Killing tensors}
Here, we will investigate the Killing tensors for the Siklos metric.
We solved (\ref{kte}), by imposing the condition that
the Killing tensor sustains the form of the metric.
We have distinguished two cases in regards to the dependencies of the metric
function $h(x,y,u)$.\\
Case {\bf a)}~The function $h(y,u)$ depends on $y$ and $u$.
The surviving components read:
\begin{equation}
\begin{array}{l}
K_{14}=K_{33}=c_{1}/x^{2}+ c_{2} / x^{4},\\[2mm]
K_{22}=c_{1}/x^{2},\\[2mm]
K_{44}=(c_{1}/x^{2}+ c_{2} / x^{4} )\,h(y,u)+ c_{3} / x^{4}.
\end{array}
\label{kt1s}
\end{equation}
The dual metric corresponding to the above Killing tensor 
is of the following form:
\begin{equation}
\begin{array}{l}
k_{14} = k_{33}=\beta^{4}/ (c_{1} x^{2} + c_{2}), \\[2mm]
k_{22} = \beta^{4}/c_{1}x^{2}, \\[2mm]
k_{44} = \beta^{4}((c_{1} x^{2}+c_{2}) h(y,u) - c_{3})/
                  (c_{1}x^{2}+c_{2})^{2}.
\end{array}
\label{ds1}
\end{equation}
We observe that, when $c_{2}=c_{3}=0$, we obtain the
original metric, with $h$ independent of $x$.\\
Case {\bf b)}~Here the metric function $h(x)$ is only a function of $x$.
The solutions for the components of the Killing tensor are
the same as above except for $K_{44}$ which is:
\begin{equation}
K_{44}=(\frac{c_{1}}{x^{2}}+\frac{2\,c_{2}}{x^{4}})\,h(x)
+\frac{c_{3}}{x^{4}}.
\label{kt2s}
\end{equation}
For case b)  the dual metric is
\begin{equation}
k_{44} = \frac{\beta^{4}(c_{1} x^{2} h(x) - c_{3})}
{(c_{1}x^{2} + c_{2})^{2}}.
\label{ds2}
\end{equation}
Similarly, when $c_{2}=c_{3}=0$, we obtain the
original metric, with $h(x)$.

From (\ref{ky2s}) the components of the Killing tensors are
obtained as:
\begin{equation}
\begin{array}{l}
K_{14}=x^{8}\,(c\,y - 2 r(u))^{2}/4 \beta^{2}, \qquad  
K_{22}=K_{33}=-c^2\,x^8/4 \beta^{2},\\[2mm]
K_{24}= -x^2 (c\,(y\,r(u)_{,u}+ s(u) + c\,v) 
       + x\,(c\,y-2r(u))\,r(u)_{,u})/2 \beta^{2} ,\\[2mm]
K_{34}=-x^2 ((y\,r(u)_{u} + s+ c v))(c\,y - 2r)- cx\,r(u)_{u})
/2 \beta^{2},  \\[2mm]
K_{44}=(x^{2}(x^{10}(cy-2r(u))^2\,h(x,y,u)-4 r(u)_{,u}^{2})
-4(y\,r(u)_{,u}+s(u)+c\,v)^{2})/4 \beta^{2} x^{4}, 
\end{array}
\label{kfy2s}
\end{equation}
where $r(u), s(u)$ and $h(x,y,u)$ are as in (\ref{rs}) 
and (\ref{hs}).  The dual metric for this Killing tensor
is calculated, but it turned out to be somewhat cumbersome with a very
remote chance of having any interpretation.

From (\ref{kfky2}) and (\ref{ky3s}), the Killing tensors are:
\begin{equation}
\begin{array}{l}
K_{14}=-2((y+r(u))^{2}+x^{2})/\beta^{4}x^{4},\\[2mm]
K_{22}=-2/\beta^{2}x^{4},\\[2mm]
K_{23}=-2\,(r(u)+y)/\beta^{2}x^{5},\\[2mm]
K_{24}=-2\,r(u)_{,u}(r(u)+y)/\beta^{4}x^{3},\\[2mm]
K_{33}=-2\,(r(u)+y)^{2}/\beta^{2}x^{6},\\[2mm]
K_{34}=2r(u)_{,u}/\beta^{4}x^{2},\\[2mm]
K_{44}=-2  \{
\beta^{2}\,h(x,y,u) [ (r(u)+y)^{2}+x^{2} ]+x^{4}r(u)_{,u}^{2}
           \} / \beta^{4} x^{6}.
\end{array}
\end{equation}
This tensor is also degenerate, as the other tensors obtained from third
rank KY tensors.

\section{Concluding Remarks}
In this article we have solved the KY equations, and we have found
valence two and valence three KY tensors, both for the pp-wave and
the Siklos metric.  In fact, the system of equations (\ref{kye})
are complicated, involving 24 coupled equations to be solved for
six component KY tensors of valence two.  Similarly,
one has to solve for 16 coupled equations for four component KY 
tensors of valence three.  
In order to find analytical and non-trivial
solutions we have looked for non-flat subspaces
admitting the maximum number of non-zero KY components,
with minimum restrictions on the metric functions.
By analyzing the integrability conditions, we have found that
there can be at most four and at most three non-zero components
for valence two and valence three KY tensors, respectively.
We have explicitly given the relations between
the metric functions and the functions appearing in the KY tensors
and thereby we have classified the pp-wave and the Siklos metric 
admitting KY tensors.   

As to our knowledge a valence three KY tensor for a curved spacetime 
have not been exemplified yet.
Therefore, we have attained further non-generic supercharges 
as well as the usual ones obtained from valence two KY tensors.  
Moreover, valence three KY tensors can be considered as a particular 
kind of Lax tensors on the manifold.  We also note that KY
tensors with maximum number of non-zero components are all non-trivial.

In general, for an arbitrary metric, one cannot predict in advance that 
the Killing tensor equations admit non-degenerate and non-trivial
solutions, because there is not a well defined technique to solve this
problem.  For this purpose we have analyzed in detail equation (\ref{kte}) 
and looked for non-degenerate Killing tensors that are of the same form as 
that of the initial Siklos metric.
Furthermore,  we have obtained second rank Killing tensors 
from valence two and three KY tensors.  
We have noted that,
the Killing tensors obtained by solving the Killing equation
are quite different, from those obtained by contracting
the KY tensors.  It is well known that, irrespective of how
they are obtained, Killing tensors determine the first integrals,
and play an important role in the solution of the geodesic equation.  
We have found that the Killing tensors obtained from the second rank KY
tensors give non-degenerate and non-trivial Killing tensors, which can be
considered as dual metrics.
On the other hand, we have observed that Killing tensors obtained from
valence three KY tensors yielded degenerate Killing tensors that are not
suitable for metric generation.     
We have investigated the dual metrics for possible physical
interpretations and we have found that the dual metric obtained from the
pp-wave  Killing tensor falls into a more general class of metrics
describing parallel null one-planes.

\section*{Acknowledgments}
\noindent
We would like to thank to J. Podolsky for valuable suggestions and
encouragements,
and C. D. Collinson for providing us with some of the crucial references.

\end{document}